\documentstyle[aps,prl,preprint,floats,epsfig]{revtex}

\newcommand{\dstlnu}{\mbox{$D^*\ell\nu$}}

\newcommand{\dstplnu}{\mbox{$D^{*+}\ell\nu$}}

\newcommand{\dstp}{\mbox{$D^{*+}$}}
    
\newcommand{\dstxlnu}{\mbox{$D^* X\ell\nu$}}    
\newcommand{\vcb}{\mbox{$|V_{cb}|$}}

\newcommand{\bbbar}{B\bar{B}}

\newcommand{\deltam}{\mbox{$\Delta m$}}

\newcommand{\cby}{\mbox{$\cos\theta_{B-D^*\ell}$}}
\newcommand{\gev}{\ \rm GeV}
\newcommand{\mev}{\ \rm MeV}
\newcommand{\invps}{\mbox{\ ${\rm ps}^{-1}$}}
\newcommand{\invfb}{\mbox{\ ${\rm fb}^{-1}$}}
\newcommand{\etal}{{\it et al.}}

\begin{document}

\preprint{\tighten\vbox{\hbox{\hfil CLEO CONF 00-03}
                        \hbox{\hfil ICHEP00 770}
}}

\title{Determination of the $B \rightarrow D^*\ell\nu$ 
Decay Width and $|V_{cb}|$}  

\author{CLEO Collaboration}
\date{\today}

\maketitle
\tighten

\begin{abstract} 

We determine the CKM matrix element \vcb\ using a sample of
3.33 million $B\bar{B}$ events in the CLEO detector at CESR. 
We determine the yield of 
reconstructed $B\to\dstplnu$ decays as a
function of $w = v_B \cdot v_{D^*}$, and from this we obtain
the differential decay rate $d\Gamma/dw$.  By extrapolating 
$d\Gamma/dw$ to $w=1$, 
the kinematic 
point at which the $D^*$ is at rest relative 
to the $B$, we extract the product $\vcb F(1)$, where $F(1)$ is
the form factor at $w=1$ and is predicted accurately by theory.  
We find $F(1)\vcb = 0.0424\pm0.0018{\rm (stat.)}\pm 0.0019{\rm (syst.)}$.
We also integrate the differential decay rate over $w$ to
obtain ${\cal B}(B\to \dstplnu) = (5.66 \pm 0.29 \pm 0.33)\%$.
All results are preliminary.

\end{abstract}
\vfill
\begin{flushleft}
.\dotfill .
\end{flushleft}
\begin{center}
Submitted to XXXth International Conference on High Energy Physics, July
2000, Osaka, Japan
\end{center}

\newpage

{
\renewcommand{\thefootnote}{\fnsymbol{footnote}}
\begin{center}
J.~P.~Alexander,$^{1}$ R.~Baker,$^{1}$ C.~Bebek,$^{1}$
B.~E.~Berger,$^{1}$ K.~Berkelman,$^{1}$ F.~Blanc,$^{1}$
V.~Boisvert,$^{1}$ D.~G.~Cassel,$^{1}$ M.~Dickson,$^{1}$
P.~S.~Drell,$^{1}$
% ...
\end{center}
 
\small
\begin{center}
$^{1}${Cornell University, Ithaca, New York 14853}\\
% ...
\end{center}
\setcounter{footnote}{0}
}
\newpage

\section{Introduction}
The CKM matrix element \vcb\ sets the length of the base of
the famous unitarity triangle.  One strategy for determining \vcb\
uses the decay $B\to\dstlnu$.  The rate for this decay, however,
depends not only on \vcb\ and well-known weak decay physics, but also on
 strong interaction effects, which are parametrized
by form factors.  These effects are notoriously difficult
to quantify, but Heavy Quark Effective Theory (HQET) offers
a method for calculating them at the kinematic point at which
the final state $D^*$ is at rest with respect to the initial
$B$ meson ($w = v_{D^*}\cdot v_B = 1$, and is the relativistic
boost $\gamma$ of the $D^*$ in the $B$ rest frame).  In this analysis, we
take advantage of this information: we measure $d\Gamma/dw$ for 
these decays, and extrapolate to obtain the rate at $w=1$.  The
rate at this point is
proportional to $[\vcb F(1)]^2$ where $F(w)$ is the form
factor.
Combined with the theoretical results, this gives \vcb.

The analysis uses $B^0\to\dstplnu$ decays and their charge conjugates
(charge conjugates are implied throughout this paper).
 We divide the reconstructed events into bins of $w$. In each
bin we extract the yield of \dstlnu\ decays using a fit
to the distribution \cby, where
\begin{equation}
\cby = \frac{2E_B E_{D^*\ell} - m_B^2 - m_{D^*\ell}^2}{2|{\bf p}_B||{\bf p}_{D^*\ell}|}.
\label{eq:cby}
\end{equation}
The angle \cby\ is thus the reconstructed angle between the $D^*$-lepton 
combination 
and the $B$ meson, computed with the assumption that the only missing mass is that of the 
neutrino.  This distribution distinguishes $B\to\dstlnu$
decays from decays such as $B\to D^{**}\ell\nu$, since \dstlnu\ decays
are concentrated in the physical
region, $-1 \le \cby <1$, while the larger missing mass
of the $D^{**}\ell\nu$ decays allows them to populate $\cby < -1$.
Given the \dstlnu\ yields as a function of $w$, we fit for a parameter
describing the form factor and the normalization at $w=1$.
  This normalization is 
proportional to the product $[\vcb F(1)]^2$.

\section{Event Samples}
We do our analysis with
3.33 million $\bbbar$ events (3.1 \invfb) 
produced on the $\Upsilon(4S)$ resonance at the Cornell Electron
Storage Ring and detected in the CLEO II 
detector~\cite{cleonim}.
In addition, the analysis uses a sample of 1.6 \invfb\ of data collected
slightly below the $\Upsilon(4S)$ resonance for the purpose of subtracting
continuum backgrounds.

The analysis uses events from a GEANT-based~\cite{geant} Monte Carlo simulation
to provide the distribution of \dstlnu\ events
in \cby\ and to provide information on some backgrounds.
 In the simulation, 
\dstlnu\ decays are modeled using a linear form factor (for $h_{A_1}(w)$)
with the parameters 
measured in a previous CLEO analysis~\cite{ffprl}.  The signal
includes events with
final state radiation as modeled by PHOTOS~\cite{photos}. We simulate
other
form factors by reweighting this sample.
Non-resonant $B\to D^*\pi\ell\nu$ decays are modeled 
using the results of Goity and Roberts~\cite{goityroberts}, and 
$B\to D^{**}\ell\nu$ 
decays are modeled using the ISGW2~\cite{isgw2} form factors.
In the following, we refer to $B\to D^{**}\ell\nu$ and 
non-resonant $B\to D^*\pi\ell\nu$ collectively as \dstxlnu\ decays.

\section{Event Reconstruction}
We use 
the decay chain $D^{*+} \to D^0 \pi^+$ followed by 
$D^0\to K^-\pi^+$. 
We first combine kaon and pion candidates in hadronic events 
 to form $D^0$ candidates. 
Signal events lie in the window 
$|m_{K\pi}-1.865| \le 0.020$\gev.  We then add a slow $\pi^+$ 
to the $D^0$ candidate to form a \dstp.  
The $K$ and $\pi$ are fit to a common vertex
and then the slow $\pi$ and $D$ are fit to a second
vertex using the beam spot constraint. This second vertex
constraint improves the resolution in 
$\deltam = m_{K\pi\pi}- m_{K\pi}$ by about 20\%.
We require $|\deltam - 0.14544|\le 0.002$ \gev.

Electrons are identified using the ratio of their energy deposition in
the CsI calorimeter to the reconstructed track momentum, the shape
of the shower in the calorimeter, and their specific ionization in the
drift chamber.  Our candidates lie in the momentum range $0.8~<~p_e~\le~2.4$~\gev.
Muon candidates penetrate two layers of steel in the solenoid return
yoke ($\approx 5$ interaction lengths). Only muons above about 
1.4 \gev\ satisfy this requirement; we therefore demand that they lie in the momentum range $1.4 < 
p_\mu \le 2.4$\gev. The charge of 
the lepton must match the charge of the kaon, and
be opposite that of the slow pion. 

Exact reconstruction of $w$ requires knowledge of the flight direction of 
the $B$ meson.  While this is unknown, our knowledge of \cby\ limits it
 relative to that of the $D^*-\ell$ combination. We therefore compute $w$
using the directions at each end of the range and we then take the 
average.  The typical resolution in $w$ is 0.03.
   We divide our sample 
into 10 equal bins from 1.0 to 1.51, where the upper bound is just beyond
the kinematic limit of 1.504.
In a few events, the reconstructed $w$ 
falls outside its kinematic range; we assign these to the first or last bin as appropriate.
In the high $w$ bins, we suppress background with no loss of signal
efficiency by restricting the angle between the $D^*$ and the lepton.

\section{Extracting the \dstlnu\ Yields}
\subsection{Method}
At this stage, our sample of candidates contains not only \dstlnu\ events, but also 
$B\to D^*X \ell\nu$ decays and various backgrounds.
In order to disentangle the \dstlnu\ from the
\dstxlnu\ decays, we 
use a binned maximum likelihood fit~\cite{likelihood} to the \cby\ 
distribution.  In this fit, the normalizations of 
the various background distributions are fixed and we allow the 
normalizations of the \dstlnu\ and the \dstxlnu\
events to float, 
with the constraint that both normalization be positive (or zero) and that 
the total event yield matches that of the data.  

The distributions of the \dstlnu\ and \dstxlnu\ decays come 
from our signal Monte Carlo. The backgrounds, and how we obtain their 
\cby\ distributions and 
normalizations, are described in the next section.

\subsection{Backgrounds}
There are several sources of events other than $B\to \dstlnu$ and 
$B\to \dstxlnu$.  
We divide these backgrounds into five classes: continuum, combinatoric,
uncorrelated, correlated and fake lepton.

\subsubsection{Continuum Background}
At the $\Upsilon(4S)$ we detect not only resonance events, but
also non-resonant events such as $e^+e^- \to q\bar{q}$. This
background contributes about 4\% of the events 
within the range $-1 <\cby 
\le 1$ (the ``signal region''). In order
to subtract background from this source, 
CESR runs one-third of the time
slightly below the $\Upsilon(4S)$ resonance.
For this continuum background, we use the \cby\ distribution of 
events in the off-resonance data scaled
by the ratio of luminosities  
and corrected for the 
small difference in the cross sections at the two center of mass energies.

\subsubsection{Combinatoric Background} 
Combinatoric background events are those in which one or more of 
the particles in the $D^*$ candidate does not 
come from a true $D^*$ decay.
This background contributes 6\% of the events in the signal region.

\begin{figure}
\centering
\epsfxsize=3.in
\epsfig{file=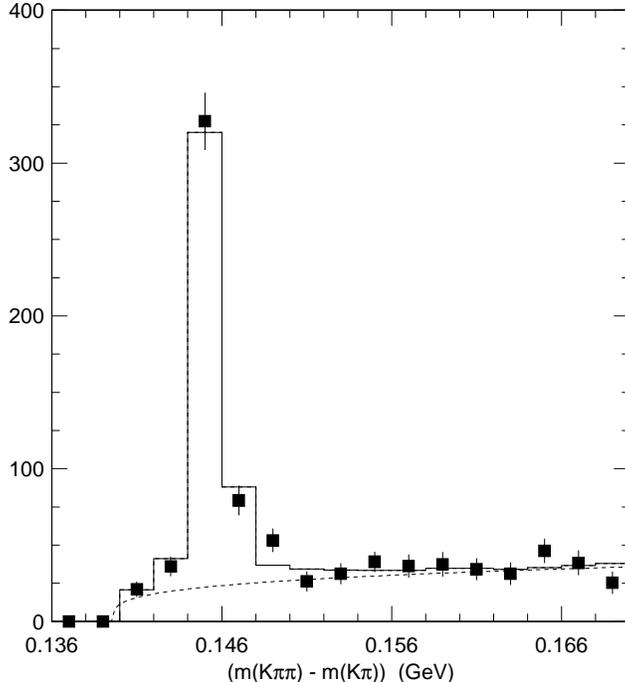,width=0.5\textwidth}
\caption{The \deltam\ distribution of events in the fifth $w$ bin, for
all \cby.
The data (solid squares) are superimposed
with the combinatoric background distribution (dashed curve) and the sum 
of the background and the \dstp\ signal (solid histogram).  
The \cby\ distribution of the
combinatoric events is taken from events in the region 
$0.155 \le \deltam < 0.165$ \gev.}
\label{fig:deltamfit}
\end{figure}

We take the \cby\ distribution of combinatoric background events
from the high \deltam\ sideband 
($0.155<\deltam\le 0.165 $\gev).  Their normalization comes from
a fit to the \deltam\ distribution in which we assume
a background shape
of the form $n(\deltam - m_\pi)^{1-c^2}$, and vary $n$, $c$,
and the normalization of the signal peak.  The lineshape for the
peak is taken from Monte Carlo.  This fit is shown for a representative
$w$ bin in Figure~\ref{fig:deltamfit}.

\subsubsection{Uncorrelated background}
Uncorrelated background, which accounts for approximately 4\% of the
events in the signal region, arises when the $D^*$ and lepton come 
from the decays of 
different $B$ mesons. 
Most of this background consists of a $D^*$ meson combined with
a secondary lepton (that is, a lepton from the chain $b\to c\to s\ell\nu$)
because primary leptons from the other $B$ have the wrong charge 
correlation. Uncorrelated background events can also arise, however, when the 
$B^0$ and $\bar{B}^0$ mix or
when a $D^*$
from the upper-vertex (that is, from the $\bar{c}$ in the decay chain $b\to c\bar{c}s$) is combined
with a primary lepton.

We obtain the \cby\ distribution of this background by 
simulating each of the various
sources and
 normalizing  each one appropriately.  To normalize, we use
the inclusive
$D^*$ production rate observed in our data, the measured
primary and secondary lepton
decay rates~\cite{roywang}, the estimated decay rate for modes in the $B\to D^{(*)} D^* K^{(*)}$ family,
and the measured $B^0 - \bar{B}^0$ mixing rate~\cite{bbmix}.
Since the \cby\ distribution depends somewhat on the momentum of the $D^*$,
we normalize the $D^*$ sources separately in low and high momentum bins.

\subsubsection{Correlated Background}
Correlated background events are those
in which the $D^*$ and lepton are daughters of 
the same $B$, but the decay was not $B\to\dstlnu$ or 
$B\to\dstxlnu$.  The most 
common sources are $B\to D^*\tau\nu$ followed by leptonic $\tau$ decay, 
and $B \to D^* D_s$ followed by semileptonic decay of the $D_s$.  
This background accounts for less than 0.5\% of the events in the signal
region and is provided by Monte Carlo simulation.

\subsubsection{Fake Lepton Background}
Fake lepton background arises when a hadron is misidentified as a lepton and is then used in our reconstruction. 
A preliminary study indicates that this background is small, and
we ignore it.

\subsection{$B\to\dstlnu$ Yields}
\begin{figure}
\centering
\hbox{
\epsfxsize=3.in
\epsfysize=3.in 
\epsfig{file=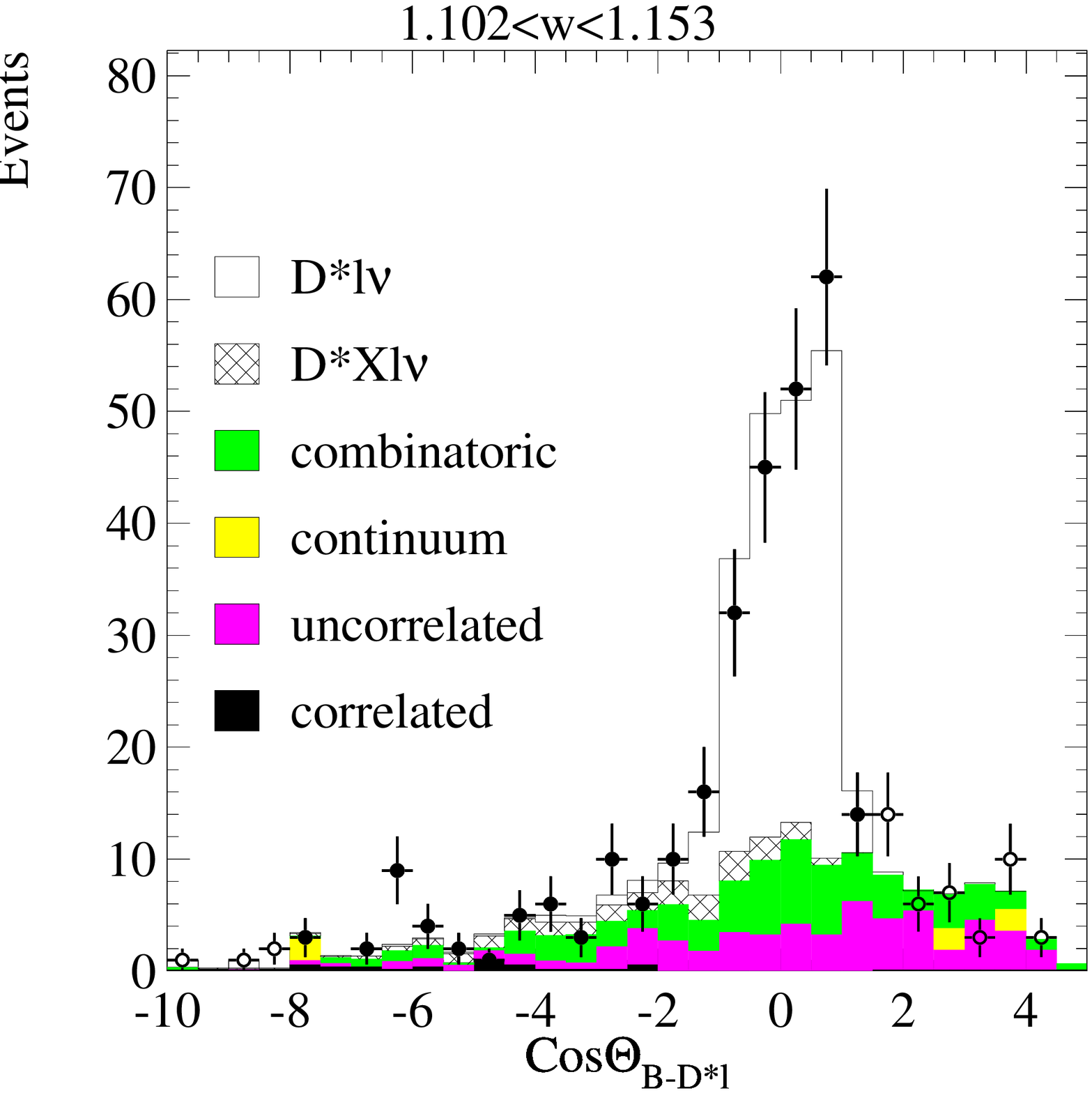,width=0.5\textwidth}
\epsfxsize=3.in
\epsfysize=3.in 
\epsfig{file=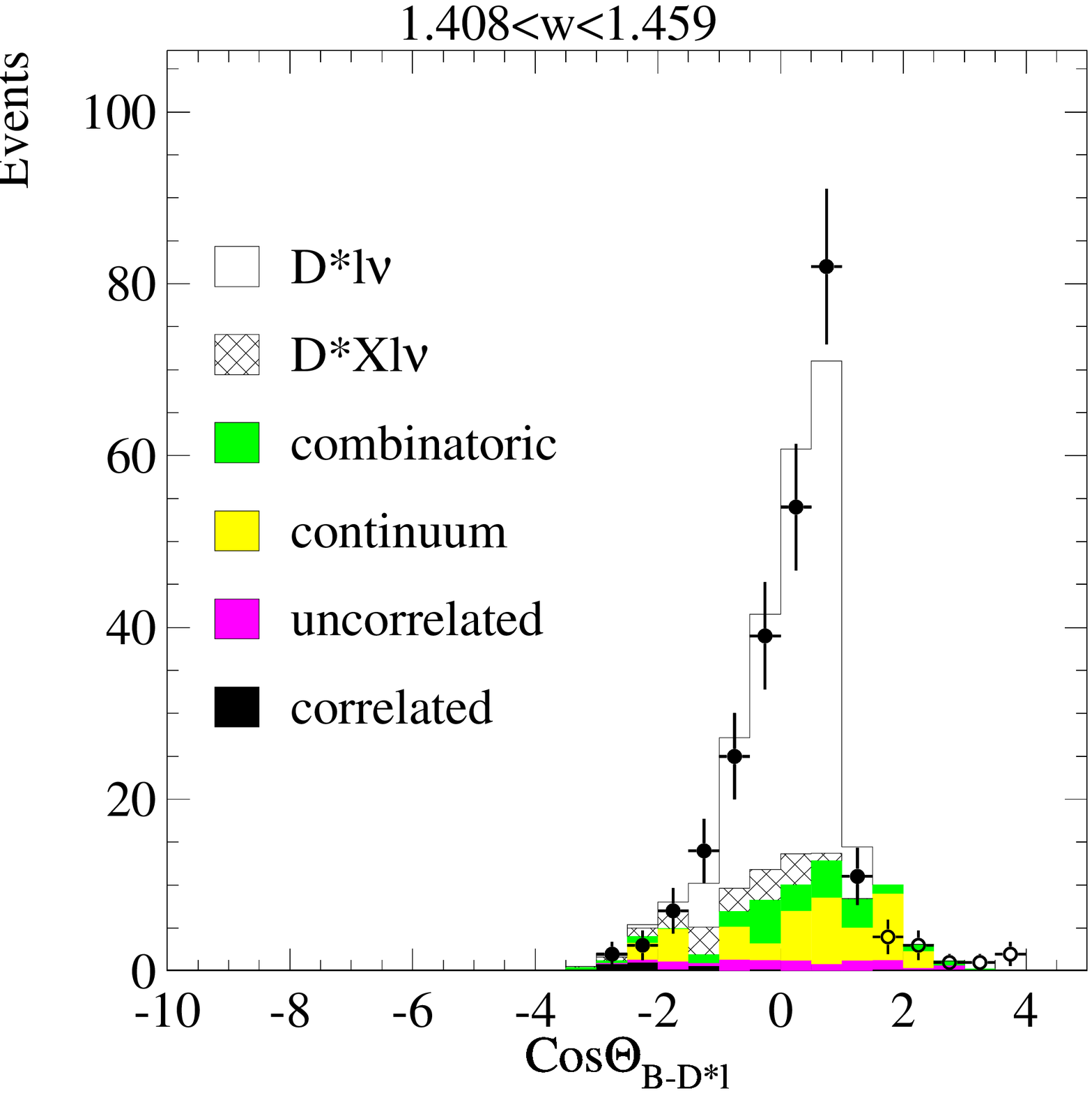,width=0.5\textwidth}
}
\caption{The event yields (circles) in the third and ninth $w$ bins with
the results of the fit superimposed.  The fit range is $-8\le\cby<1.5$,
and is indicated with the solid circles.}
\label{fig:dstpwbin39}
\end{figure}

\begin{figure}
\centering
\epsfxsize=4.in 
\epsfig{file=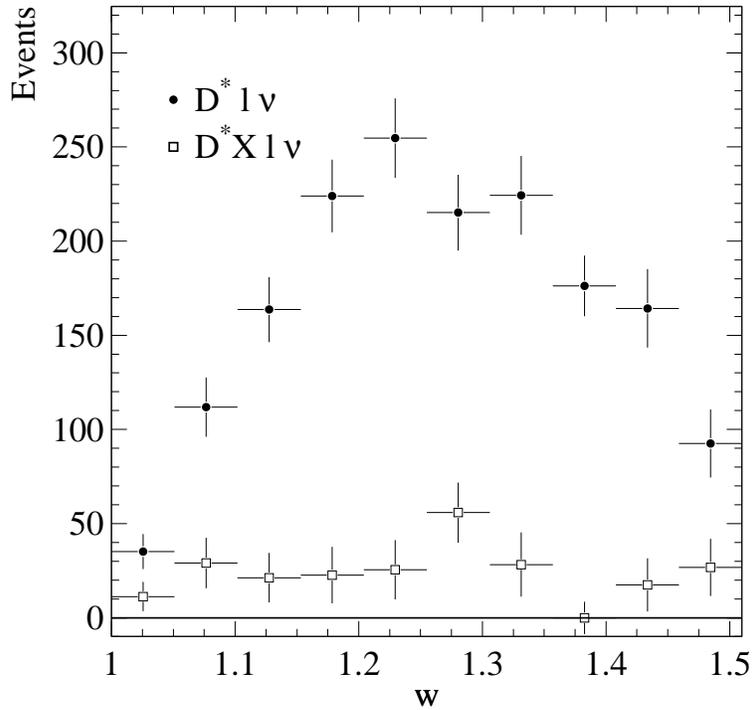,width=4in}
\caption{The \dstlnu\ and $\dstp X \ell\nu$ yields in each
$w$ bin.}
\label{fig:yields}
\end{figure}

Having obtained the distributions in \cby\ of the signal and background
components, we fit for the yield of \dstplnu\ events in 
each $w$ bin.
Two representative fits are shown in Figure~\ref{fig:dstpwbin39}.
The quality of the fits is 
good, as is agreement between the 
data and fit distributions outside the fitting region.  We summarize 
the \dstlnu\ and
\dstxlnu\ yields in Figure~\ref{fig:yields}.

\begin{figure}
\centering
\hbox{
\epsfig{file=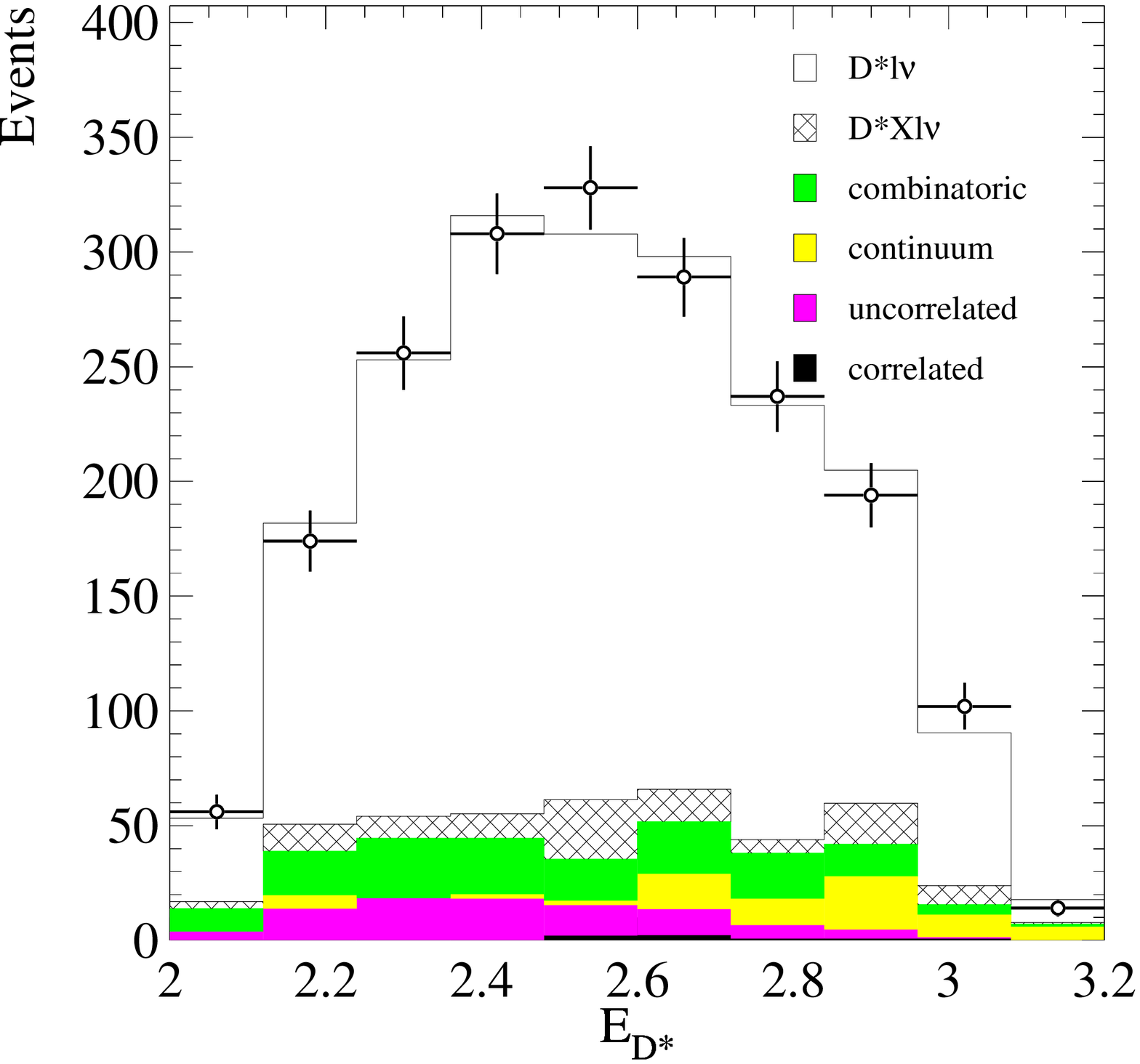,width=0.5\textwidth}
\epsfig{file=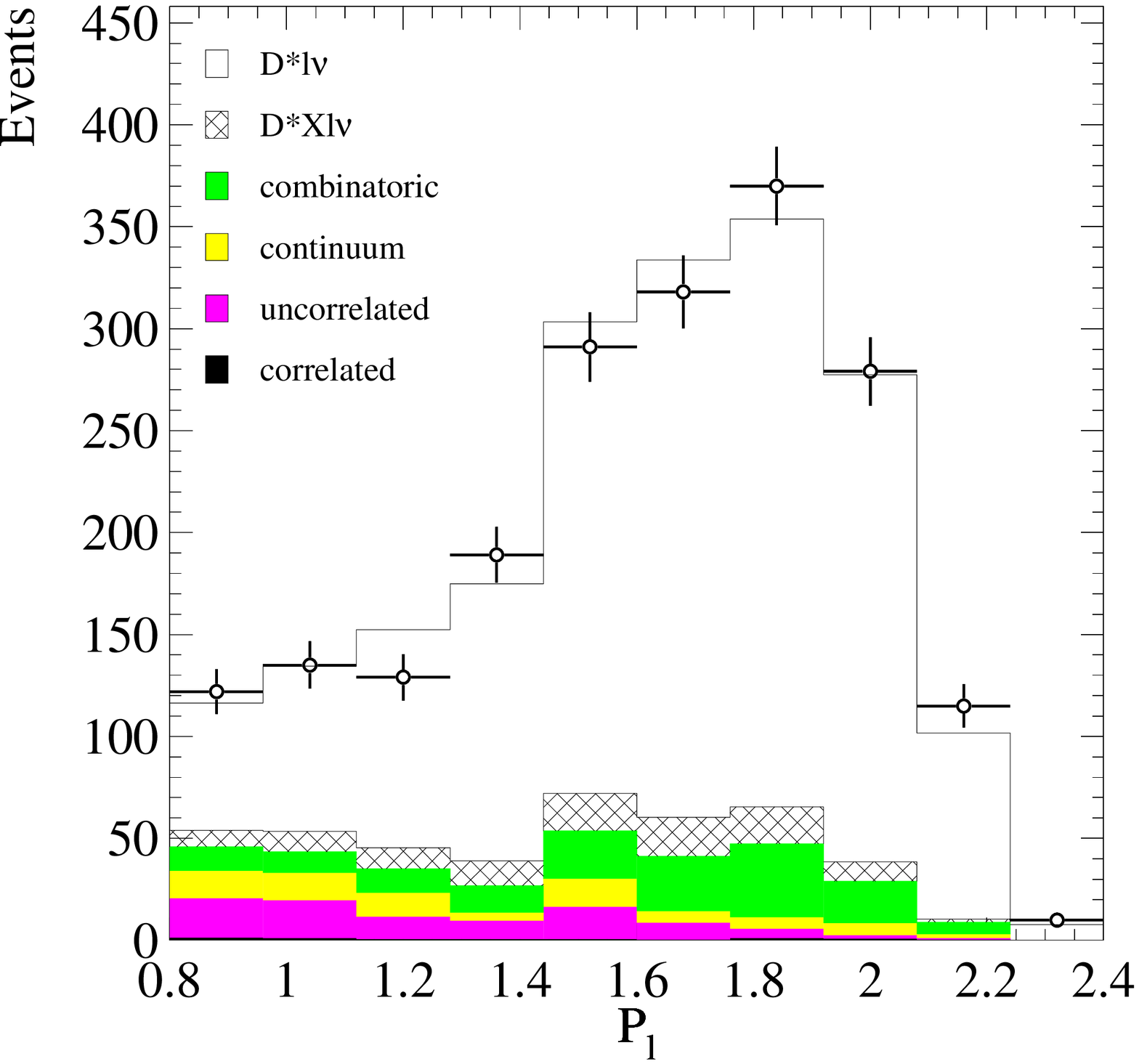,width=0.5\textwidth}
}
\caption{The energy distribution of \dstp\ candidates (left) and the
lepton momentum spectrum for \dstplnu\ candidates (right) in the
signal region for all $w$ bins combined.} 
\label{fig:edst}
\end{figure}

In order to test the quality of our fit and modeling of the signal
and backgrounds, we plot a variety of distributions.
Figure~\ref{fig:edst} shows the
$D^*$ energy distribution and
the lepton
momentum spectrum.
We find good agreement between the data and our expectations.

\section{The \vcb\ Fit}
The partial width for $B\to\dstlnu$ decays is given by~\cite{richburch}
\begin{equation}
\frac{d\Gamma}{dw}=\frac{G_F^2}{48 \pi^3}
(m_B-m_{D^*})^2 m_{D^*}^3 \sqrt{w^2-1}(w+1)^2 
\left(1 + 4\left(\frac{w}{w+1}\right)
\left(\frac{1-2wr+r^2}{(1-r)^2}\right)\right)
\left[\vcb F(w)\right]^2
\end{equation}
where $m_B$ and $m_{D^*}$ are the $B$ and $D^*$ meson masses,
$r=m_{D^*}/m_B$, and the form factor $F(w)$ is given by
\begin{equation}
F(w) = 
\sqrt{\frac{\tilde{H}^2_0 + \tilde{H}^2_+ + \tilde{H}^2_-}
{1 + 4\left(\frac{w}{w+1}\right)\left(\frac{1-2wr+r^2}{(1-r)^2}\right)}}
h_{A_1}(w).
\end{equation}
The $\tilde{H}_i$ are the helicity form factors
and are given by
\begin{eqnarray}
\tilde{H}_0(w) &=& 1 + \frac{w-1}{1-r}(1-R_2(w))\\
\tilde{H}_+(w) &=& \frac{\sqrt{1-2wr+r^2}}{1-r}\left( 1-\sqrt{\frac{w-1}{w+1}}R_1(w)\right)\\
\tilde{H}_-(w) &=& \frac{\sqrt{1-2wr+r^2}}{1-r}\left( 1+\sqrt{\frac{w-1}{w+1}}R_1(w)\right).
\end{eqnarray}
The form factor $h_{A_1}(w)$ and the form factor ratios
$R_1(w) = h_V(w)/h_{A_1}(w)$ and $R_2(w) =(h_{A_3}(w)+rh_{A_2}(w))/h_{A_1}(w)$ have been studied both experimentally
and theoretically.  A CLEO analysis~\cite{ffprl} measured these form
factors under the assumptions that $h_{A_1}(w)$
is linear as a function of $w$ and that $R_1$ and
$R_2$ are independent of $w$.  CLEO found 
\begin{eqnarray}
R_1 &=& 1.18 \pm 0.30 \pm 0.12, \\
R_2 &=& 0.71 \pm 0.22 \pm 0.07 {\rm \ and} \\
-\frac{dh_{A_1}}{dw}(w=1)\equiv \rho^2 &=& 0.91 \pm 0.15 \pm 0.06
\end{eqnarray}
with the correlation coefficients $C(\rho^2,R_1)=0.60$, $C(\rho^2,R_2) = -0.80$ and $C(R_1, R_2) =-0.82$. 

$R_1(1)$ and $R_2(1)$ have been computed using HQET and QCD sum rules
with the results
$R_1(1)=1.27$ and $R_2(1)=0.8$ and estimated errors of
$0.1$ and $0.2$ respectively~\cite{neubert}, in good agreement
with the experimental results. $R_1(w)$ and $R_2(w)$ 
are expected to vary weakly with $w$.
Most importantly for this analysis,
$F(1)(=h_{A_1}(1))$ is relatively well-known, thereby allowing us
to disentangle it from \vcb.  We will use 
$F(1)=0.913\pm 0.042$~\cite{babarbook}.

Recently, dispersion relations have been used to constrain the
shapes of the form factors\cite{caprini},~\cite{lebed}. Rather than
expand the form factor in $w$, these analyses
expand in the variable $z=(\sqrt{w+1}-\sqrt{2})/(\sqrt{w+1}+\sqrt{2})$. 
Ref.~\cite{caprini} obtains the results:
\begin{eqnarray}
\label{eq:formfact1}
h_{A_1}(w) &=& 1 - 8\rho^2z +
              ( 53\rho^2-15 )z^2 -
              ( 231\rho^2 - 91 )z^3 \\
\label{eq:formfact2}
R_1(w) &=& R_1(1) - 0.12(w-1) + 0.05(w-1)^2\\
R_2(w) &=& R_2(1) + 0.11(w-1) - 0.06(w-1)^2.
\label{eq:formfact3}
\end{eqnarray}

In our analysis, we assume that the form factor has the
functional form derived from dispersion relations given
in Equations~\ref{eq:formfact1},~\ref{eq:formfact2} and~\ref{eq:formfact3}.  We
fit our yields as a function of $w$ for $F(1)$\vcb\
and $\rho^2$, keeping $R_1(1)$ and $R_2(1)$ fixed at their measured values.
Our fit minimizes 
\begin{equation}
\chi^2 = \sum_{i=1}^{10}
\frac{[N_i^{obs} - \sum_{j=1}^{10}\epsilon_{ij}N_j]^2}
{\sigma_{N_i^{obs}}^2},
\end{equation}
where $N_i^{obs}$ is the yield in the $i^{{\rm th}}$ $w$ bin,
$N_j$ is the number of decays in the $j^{{\rm th}}$ $w$ bin, and 
the matrix $\epsilon_{ij}$ accounts for the reconstruction efficiency
and the smearing in $w$.
Explicitly,
\begin{equation} 
N_j = 4 f_{00} N_{\Upsilon (4S)}{\cal B}(\dstp\to D^0\pi^+) {\cal B}(D^0\to K^-\pi^+)\tau_{B^0}\int_{w_j}dw  
\frac{d\Gamma}{dw}
\end{equation}
where $\tau_{B^0}$ is the $B^0$ lifetime~\cite{pdg}, 
${\cal B}(\dstp\to D^0\pi^+)$ is the $\dstp \to D^0\pi^+$ branching
fraction~\cite{pdg},
${\cal B}(D\to K\pi)$ is the $D^0 \to K^-\pi^+$ branching 
fraction~\cite{pdg}, $N_{\Upsilon (4S)}$ is the number of 
$\Upsilon (4S)$ events in the sample, and 
$f_{00}$ represents the
$\Upsilon (4S)\to B^0 \bar{B}^0$ branching fraction~\cite{sylvia}.
We use the result of Ref.~\cite{sylvia} for $f_{00}$ as  
a constraint in the fit.

\begin{figure}
\centering
\epsfxsize=4.in 
\epsfysize=6.in 
\epsfig{file=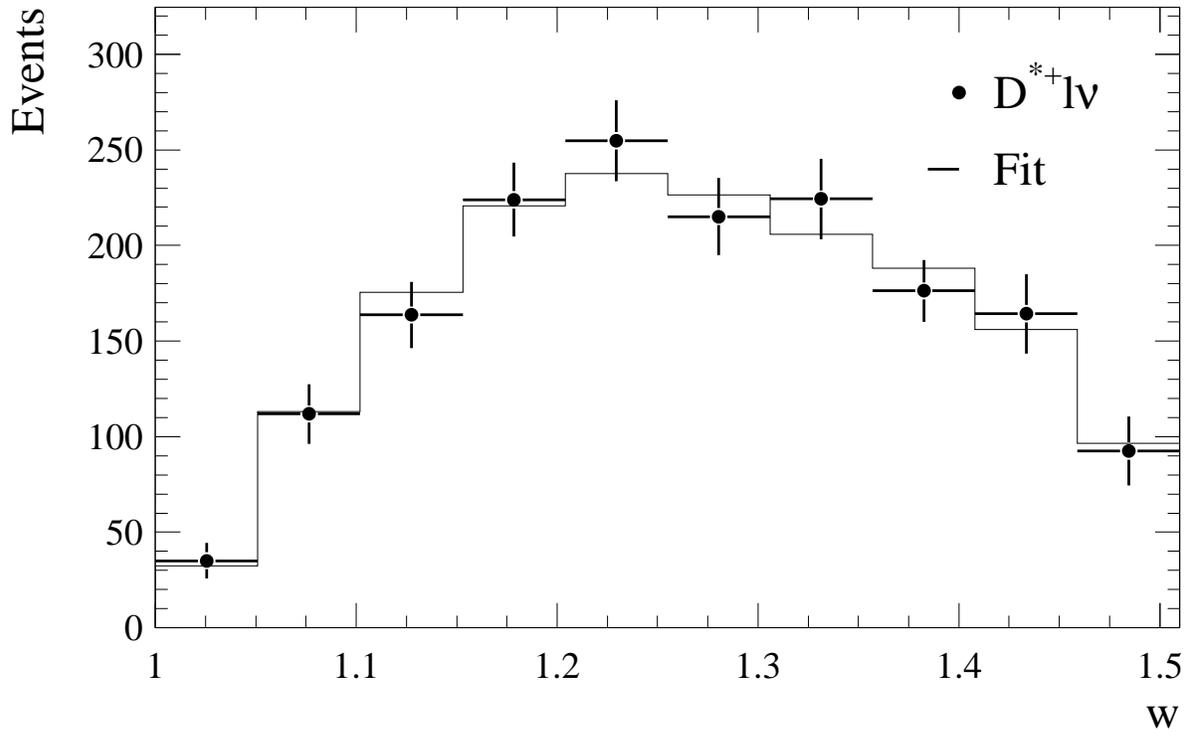,width=\textwidth}
\epsfig{file=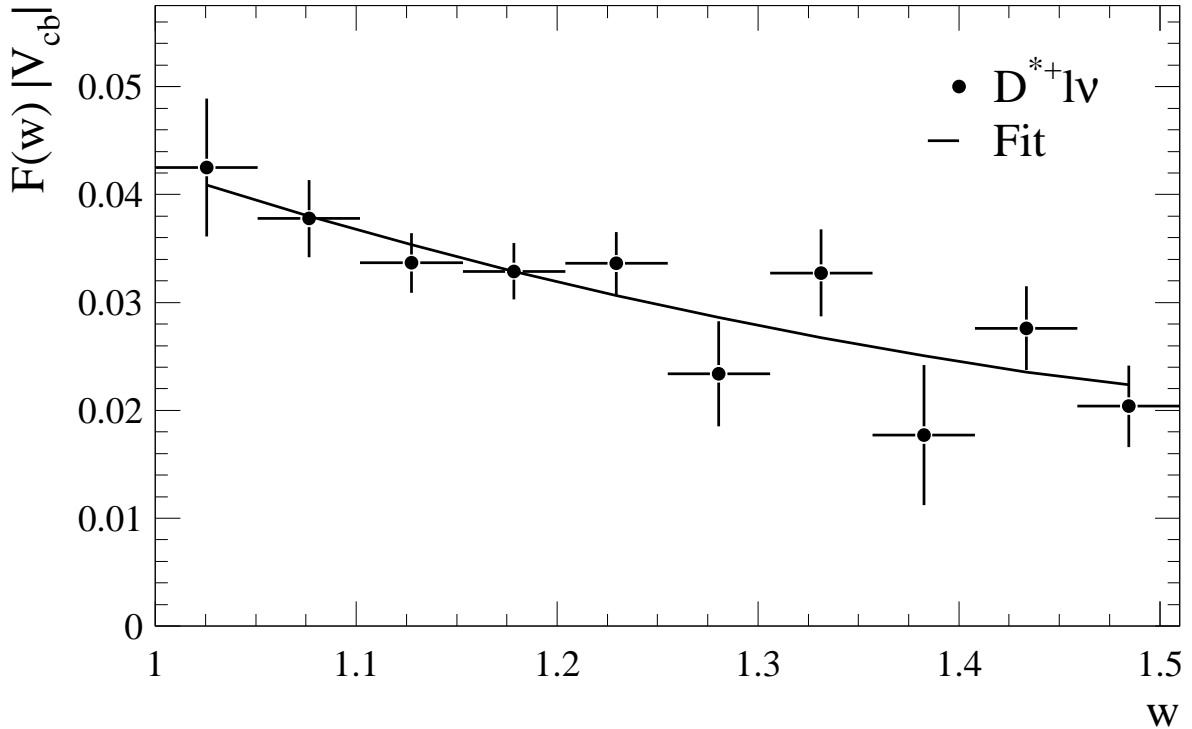,width=\textwidth}
\caption{The results of the fit to the $w$ distribution.
The upper figure shows the yields (solid circles) with the 
results of the fit
superimposed (histogram).  The lower figure displays $\vcb F(w)$, where the
data points (solid circles) are derived from the yields after correcting
for efficiency, smearing, and all terms in the differential
decay rate apart from $\vcb F(w)$.
The curve shows the result of the fit.}
\label{fig:fffit}
\end{figure}

The result of the fit is shown in Figure~\ref{fig:fffit}.
We find
\begin{eqnarray}
|V_{cb}|F(1) &=& 0.0424 \pm 0.0018\\
\rho^2 &=& 1.67 \pm 0.11 {\rm\ and}\\
\chi^2 &=& 3.1/8 {\rm\ dof.} 
\end{eqnarray}
with a correlation coefficient between $\vcb F(1)$ and $\rho^2$ of 0.90.
These parameters give $\Gamma = 0.0366\pm 0.0018 \invps$.
The quality of the fit is excellent.  We note that the slope is 
higher than that found in the previous CLEO analysis~\cite{oldcleo}
because of the curvature introduced
into our form factor. If we use a linear form factor and the same
subset of the data, we obtain results
compatible with the earlier analysis.

\section{Systematic Uncertainties}
The systematic uncertainties 
are summarized in Table~\ref{tab:syserrdstp}.
The dominant systematic uncertainties arise from our background
estimations and from our knowledge of the slow pion reconstruction efficiency.

\subsection{Background uncertainties}
 We test our procedure for subtracting combinatoric background
by applying it to Monte Carlo simulated events.  
We assign a systematic error based on the difference between 
the results obtained using the ``true'' background and those
obtained using the same procedure that we apply to our data.
We also include the statistical uncertainty of the study.

The main source of uncertainty from the uncorrelated background
is the normalization of the various contributions.  Of these, 
the most important
is the branching fraction of the $B\to D^{(*)}D^*K^{(*)}$ decays, which we vary by 50\%.
Smaller effects arise from the
primary and secondary lepton rates and from the uncertainty
in $B^0-\bar{B}^0$ mixing.

We assess the uncertainty arising from the correlated background
by varying the branching fractions of the contributing modes.

\subsection{Slow $\pi$ reconstruction uncertainty}
A major source of uncertainty for the analysis is the efficiency
for reconstructing the slow pion from the $D^*$ decay. 
  This efficiency
is low near $w =1 $ and increases rapidly over the next few $w$ bins.  
We have explored the efficiency as a function
of the event environment and as a function of hit resolution, hit
efficiency, material, and the charge division resolution in one
of the inner drift chambers.  The last of these is important
because very soft tracks often make use of the charge division
information for reconstruction of the $z$ component (beam direction) of their
trajectory. 
The uncertainty in \vcb$F(1)$ and the decay width are
dominated by uncertainties in the amount of material in the inner
detector (2.3\%) and  the drift chamber hit efficiency (0.8\%).


\subsection{Other uncertainties}
The efficiency for identifying electrons has been evaluated
using radiative bhabha events embedded in hadronic events,
and has an uncertainty of 2.4\%. Similarly, the muon identification
efficiency has been evaluated using radiative mu-pair events,
and has an uncertainty of 1.4\%.  The total uncertainty in 
lepton identification, weighted by the electron and muon populations, is 2.1\%.
Separate electron and muon analyses of our data give consistent
results.

The $B^0$ momentum is measured directly in the data using
fully reconstructed hadronic decays, and is known on average
 with a precision of 0.0016 \mev.
Variation of the momentum in our reconstruction slightly alters the \cby\ distribution
that we expect for our signal, and it therefore changes the yields obtained from
the \cby\ fits.
Likewise, CLEO has measured the $B^0$ meson mass~\cite{ershov}
and when we vary the mass within its measurement error, we find a small effect on the yields.

We determine the tracking efficiency 
uncertainties for the lepton and the $K$ and $\pi$ forming 
the $D^0$ in the same study used for the slow pion
from the \dstp\ decay.  These uncertainties are confirmed
in a study of 1-prong versus 3-prong $\tau$ decays.

\begin{figure}
\centering
\hbox{
\epsfxsize=3.in 
\epsfig{file=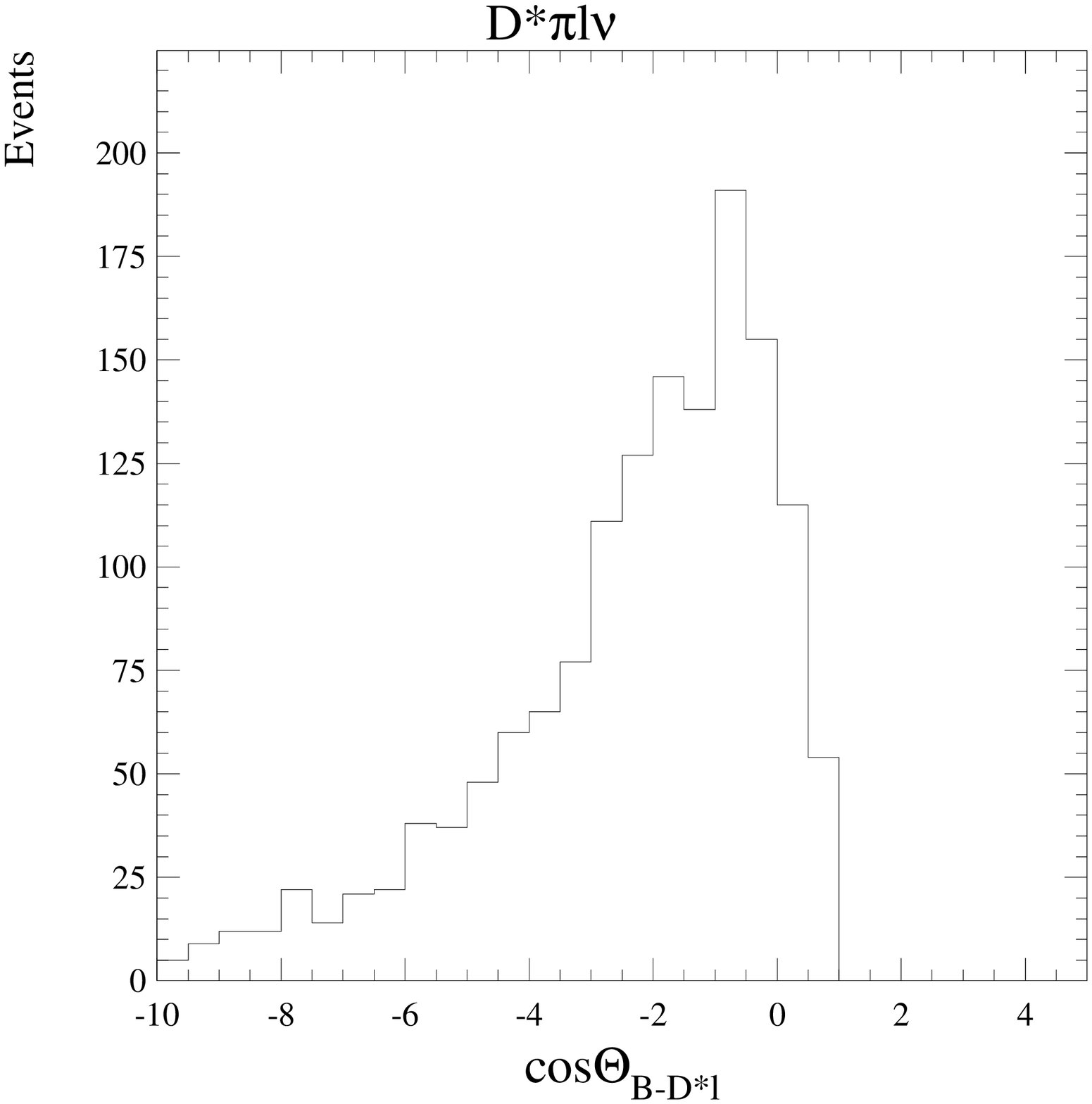,width=0.5\textwidth}
\epsfig{file=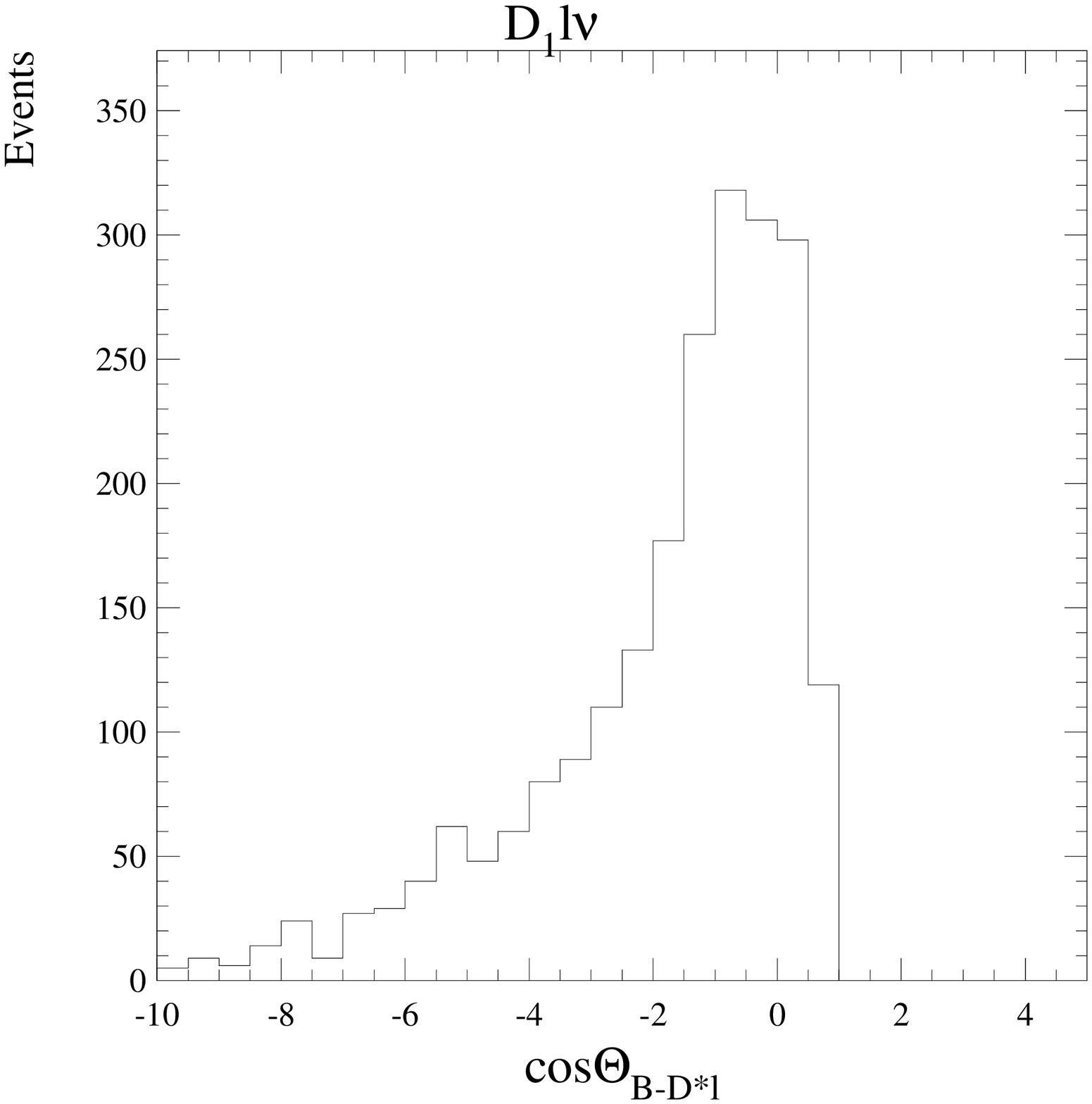,width=0.5\textwidth}
\epsfxsize=3.in 
}
\caption{The \cby\ distribution of the $B\to D^{*+}\pi\ell\nu$ (left)
and $B\to D_1\ell\nu$ (right) events contributing to the \dstplnu\
sample.}
\label{fig:ddstlnu}
\end{figure}

Finally, our analysis requires that we know the \cby\ distribution of the 
\dstxlnu\ contribution.  This distribution in turn depends on 
both the branching fractions of contributing modes and
on their form factors.  Variation of all of these branching
fractions and form factors is not only cumbersome,
but out of reach given the poor current knowledge of these modes.
Instead, we note that the $B\to D^*\pi\ell\nu$ and $B\to D_1\ell\nu$
modes are the ones with the most extreme \cby\ distributions
(the largest mean and the smallest). These distributions
are shown in Figure~\ref{fig:ddstlnu}.  
We therefore repeat the
analysis, first using pure $B\to D^*\pi\ell\nu$ to describe our
\dstxlnu\ decays and then using pure $B\to D_1\ell\nu$ to describe
these decays, and we take the larger of the two excursions
as our systematic error.

\subsection{Sensitivity to $R_1(1)$ and $R_2(1)$}
The form factor ratios
$R_1(1)$ and $R_2(1)$  affect
the lepton spectrum and therefore the fraction
of events satisfying our 0.8\gev\ electron and 1.4\gev\ muon momentum 
requirements.  To assess this effect, we vary $R(1)$ and 
$R(2)$ within their measurement
errors, taking into account the correlation between them.

\begin{table}
\centering
\caption{The fractional systematic uncertainties.}
\label{tab:syserrdstp}
\medskip
\begin{tabular}{lccc} 
Source  &  $|V_{cb}|F(1)$(\%) &$\rho^2$(\%) & $\Gamma(B \to \dstplnu)$(\%) \\ \hline
Combinatoric Background       & 1.4 & 1.8 & 1.2 \\
Uncorrelated Background       & 0.7 & 0.9 & 0.7 \\
Correlated Background         & 0.4 & 0.3 & 0.5 \\
Slow $\pi$ finding            & 3.1 & 3.7 & 2.9 \\
$K$, $\pi$ \&\ $\ell$ finding & 1.0 & 0.0 & 1.9 \\
Lepton ID                     & 1.1 & 0.0 & 2.1 \\
$B$ momentum \& mass          & 0.3 & 0.5 & 0.4 \\
\dstxlnu\ model               & 0.2 & 1.9 & 1.9 \\ 
Number of $B\bar{B}$ events   & 0.9 & 0.0 & 1.8 \\
\hline
Subtotal                      & 3.8 & 4.7 & 5.0 \\
\hline
${\cal B}(D^* \to D\pi)$      & 0.4 & 0.0 & 0.7 \\
${\cal B}(D \to K\pi)$        & 1.2 & 0.0 & 2.3 \\   
$\tau_B$                      & 1.0 & 0.0 & 2.1 \\
$R_1(1)$ and $R_2(1)$         & 1.4 &12.0 & 1.8 \\
\hline
Subtotal                      & 2.1 &12.0 & 3.7 \\
\hline
Total                         & 4.4 & 13  & 6.2 \\ 
\end{tabular}
\end{table}

\section{Conclusions}
We have fit the $w$ distribution of $B\to \dstlnu$ decays for 
the slope of the form factor and $|V_{cb}|F(1)$.
We find
\begin{eqnarray}
|V_{cb}|F(1) &=& 0.0424 \pm 0.0018\pm 0.0019\\
\rho^2 &=& 1.67 \pm 0.11 \pm 0.22\\
\end{eqnarray}
with a correlation coefficient between $\vcb F(1)$ and $\rho^2$ of 0.90.
These parameters imply the decay rate
\begin{equation}
\Gamma = 0.0366\pm 0.0018 \pm 0.0023\invps.
\end{equation}
and the branching fraction
\begin{equation}
{\cal B}(B\to \dstplnu) = (5.66 \pm 0.29 \pm 0.33)\%.
\end{equation}

Our result implies
\begin{equation}
\vcb =0.0464 \pm0.0020(stat.) \pm 0.0021(syst.)\pm 0.0021(theor.),
\end{equation}
where we have used $F(1)=0.913\pm 0.042$~\cite{babarbook}.
This is consistent with previous measurements of \vcb,
but is somewhat higher.  The analysis benefits from small
backgrounds and good resolution in $w$. 
These results are preliminary.

\section{Acknowledgements}
We are indebted to the CESR staff for the superb performance of
the accelerator.  I.P.J. Shipsey thanks the NYI program of the NSF, 
M. Selen thanks the PFF program of the NSF, 
A.H. Mahmood thanks the Texas Advanced Research Program,
M. Selen and H. Yamamoto thank the OJI program of DOE, 
M. Selen and V. Sharma 
thank the A.P. Sloan Foundation, 
M. Selen and V. Sharma thank the Research Corporation, 
F. Blanc thanks the Swiss National Science Foundation, 
and H. Schwarthoff and E. von Toerne
thank the Alexander von Humboldt Stiftung for support.  
This work was supported by the National Science Foundation, the
U.S. Department of Energy, and the Natural Sciences and Engineering Research 
Council of Canada.

\nopagebreak

\end{document}